\documentclass[onecolumn,a4paper,11pt]{cnops}
\usepackage{float}
\usepackage[breaklinks]{hyperref}
\usepackage{graphicx}
\usepackage{amsmath}
\usepackage{amssymb}

\usepackage{natbib}
\usepackage{multirow}
\bibpunct{(}{)}{;}{a}{}{,}

\newcommand{\kms}{$\rm km s ^{-1}$}

\usepackage{placeins}
\usepackage{float}

\begin{document}

\title{CNO in Giant Metal-Poor Stars }

\author{J. Alazzawi\\
\footnotesize
LIRA, Observatoire de Paris, Universit\'e PSL,Sorbonne Universit{\'e}, Universit{\'e} Paris Cit{\'e},\\
\footnotesize
 CY Cergy Paris Universit{\'e}, CNRS, 5 place Jules Janssen 92195 Meudon, France  \\
}
\maketitle

~\\
{\bf Published:} February 26, 2026 \\

\begin{abstract}
We present the chemical abundance analysis for elements carbon, nitrogen, oxygen and the carbon isotopic ratio in a sample of metal-poor stars selected from the RAVE survey based on their extreme radial velocities. Stellar parameters were derived using \textit{Gaia} photometry and parallaxes, and elemental abundances were determined from high-resolution UVES spectra obtained at the ESO Very Large Telescope. The stars span a wide metallicity range ($-2.9 \lesssim \mathrm{[Fe/H]} \lesssim -1.2$) and, for the majority of the sample, exhibit kinematic properties consistent with the Galactic outer halo population.
\end{abstract}

\section{Introduction}
Understanding how the Milky Way formed and evolved requires tracing both its chemical enrichment and its dynamical history. Kinematically hot, metal-poor stars are especially powerful probes in this respect. With their extreme 
velocities, these stars are associated with the Galactic halo and are thought to preserve signatures of the earliest phases of Galaxy formation.

Because metal-poor stars formed from a gas that has undergone little chemical enrichment, their abundance patterns provide direct insight into early nucleosynthesis. When such low metallicities are combined with extreme kinematics, the stars can also reveal past dynamical events, including accretion of satellite galaxies or strong gravitational interactions. The unprecedented astrometric precision of the \textit{Gaia} mission has transformed this field by enabling reliable kinematic selection of halo stars and efficient follow-up with high-resolution spectroscopy.

Chemical abundance ratios, particularly for elements produced on different timescales, place strong constraints on the assembly history of the Galaxy. Elements such as carbon, nitrogen and oxygen can trace the contributions of stellar populations with different masses and lifetimes.

In this work, I studied a sample of metal-poor stars selected for their high radial velocities. The targets were identified from the RAVE survey and observed with the UVES spectrograph at the ESO Very Large Telescope. The reduced UVES spectra cover the wavelength ranges 328–456 nm and 472–683 nm at a resolving power of $R \sim 40\,000$. To my knowledge, no detailed abundance analysis of these spectra has previously been published.

\section{Stellar parameters and radial velocities}

Stellar parameters were derived from \textit{Gaia} EDR3 \citep{GaiaEDR3} photometry and parallaxes. Reddening corrections were derived using the dust maps of \citet{Schlafly2011}. Effective temperatures were estimated from the $(G_{\rm BP}-G_{\rm RP})$ colour by comparison with theoretical colour--$T_{\rm eff}$ relations from the KOALA database \citep{koala}. Surface gravities were then computed via the Stefan--Boltzmann relation, using the parallax from Gaia. I applied the \textit{Gaia} parallax zero-point correction following \citet{Lindegren2021}. Starting from an initial set of atmospheric parameters, metallicities were derived using the MyGIsFOS pipeline \citep{Sbordone2014}. The final parameters are reported in Table~1. With the final stellar parameters, I computed a grid of synthetic spectra for each star, with steps in abundances of 0.2\,dex, by using the spectrum synthesis code Turbospectrum \citep{Alvarez1998, Plez2012,Plez2025}, and used these grids to drive the final [O/H] and  [Fe/H]. 

Radial velocities $V_r$ were taken from \textit{Gaia} DR3, where they are obtained from cross-correlation of the RVS spectra. Given the extreme velocities of our targets, I independently verified $V_r$ using the UVES spectra through template matching, adopting the \textit{Gaia} values as initial guesses. The comparison between \textit{Gaia} and UVES-based velocities is summarised in Table~2. Figure~1 places our sample in the broader DR3 radial-velocity distribution, for comparison with the high-velocity population discussed by \citet{Katz2025}.

\begin{table}[tb]
\centering
\caption{Final Stellar Parameters of the Sample.\label{tab:param}}
\tabcolsep=5pt
\begin{tabular*}{30pc}{@{\extracolsep\fill}lcccc@{\extracolsep\fill}}
\hline
\textbf{Star Name} & \textbf{$T_{\mathrm{eff}}$ (K)} & \textbf{$\log g$} & \textbf{$\xi$ (km\,s$^{-1}$)} & \textbf{[Fe/H]} \\
\hline
TYC 7274-00734-1  & 5032  & 2.52  & 1.54 & -1.52 \\
TYC 7535-00160-1  & 4649  & 1.43  & 1.96 & -2.67 \\
TYC 8019-00159-1  & 4864  & 1.72  & 1.92 & -2.51 \\
TYC 7524-00065-1  & 4503  & 1.17  & 1.96 & -2.04 \\
C0213360-505024   & 5486  & 2.60  & 1.72 & -1.42 \\
C2136142-694908   & 4561  & 1.80  & 1.63 & -1.19 \\
C1919566-632839   & 4069  & 0.80  & 0.79 & -1.92 \\
C1519196-191359   & 4319  & 1.34  & 1.79 & -2.08 \\
C1032508-244851   & 4970  & 2.21  & 1.69 & -2.10 \\
C1536201-144228   & 4888  & 2.22  & 1.66 & -1.96 \\
C1302091-323721   & 4223  & 0.78  & 1.98 & -1.67 \\
C1250398-030748   & 5443  & 3.20  & 1.47 & -2.15 \\
C1126364-293415   & 4196  & 1.12  & 1.79 & -1.43 \\
C1012254-203007   & 4520  & 0.73  & 2.25 & -2.91 \\
C1304064-331913   & 4189  & 1.00  & 1.85 & -1.68 \\
C1330590-230955   & 4000  & 0.24  & 2.14 & -1.85 \\
\hline
\end{tabular*}
\end{table}

\begin{table}[tb]
\centering
\caption{Radial velocities from Gaia and this work.\label{table2}}
\tabcolsep=5pt
\begin{tabular*}{30pc}{@{\extracolsep\fill}lcccc@{\extracolsep\fill}}
\hline
\textbf{Star Name} &
\textbf{$V_r^{\rm Gaia}$} &
\textbf{$\sigma V_r^{\rm Gaia}$} &
\textbf{$V_r^{\rm meas}$} &
\textbf{$\sigma V_r^{\rm meas}$} \\
 & [\kms] & [\kms] & [\kms] & [\kms] \\
\hline
TYC 7274-00734-1  & 407.9552  & 0.2078 & 408.317 & 0.4150 \\
TYC 7535-00160-1  & 355.0691  & 0.2531 & 353.523 & 0.6126 \\
TYC 8019-00159-1  & 98.4503   & 0.1710 & 99.117  & 0.7179 \\
TYC 7524-00065-1  & 350.9533  & 0.2167 & 351.920 & 0.4733 \\
C0213360-505024   & 388.2223  & 0.6106 & 388.600 & 0.7119 \\
C2136142-694908   & 382.8817  & 0.3296 & 381.640 & 0.4129 \\
C1919566-632839   & 330.3995  & 0.5303 & 328.680 & 0.5378 \\
C1519196-191359   & -409.6939 & 6.5699 & -409.610 & 0.4705 \\
C1032508-244851   & 515.3541  & 2.4702 & 509.490 & 0.4929 \\
C1536201-144228   & -328.6475 & 2.1124 & -329.600 & 0.5932 \\
C1302091-323721   & 409.8165  & 3.0273 & 410.000 & 0.4334 \\
C1250398-030748   & 396.2207  & 0.9787 & 403.020 & 0.6024 \\
C1126364-293415   & 409.6741  & 0.9257 & 409.870 & 0.4142 \\
C1012254-203007   & 508.1379  & 1.0679 & 506.755 & 0.7504 \\
C1304064-331913   & 357.8247  & 0.9944 & 360.720 & 0.4207 \\
C1330590-230955   & 334.9188  & 0.5257 & 335.350 & 0.4639 \\
\hline
\end{tabular*}
\end{table}

\begin{figure}
\centering
\includegraphics[width=0.70\textwidth]{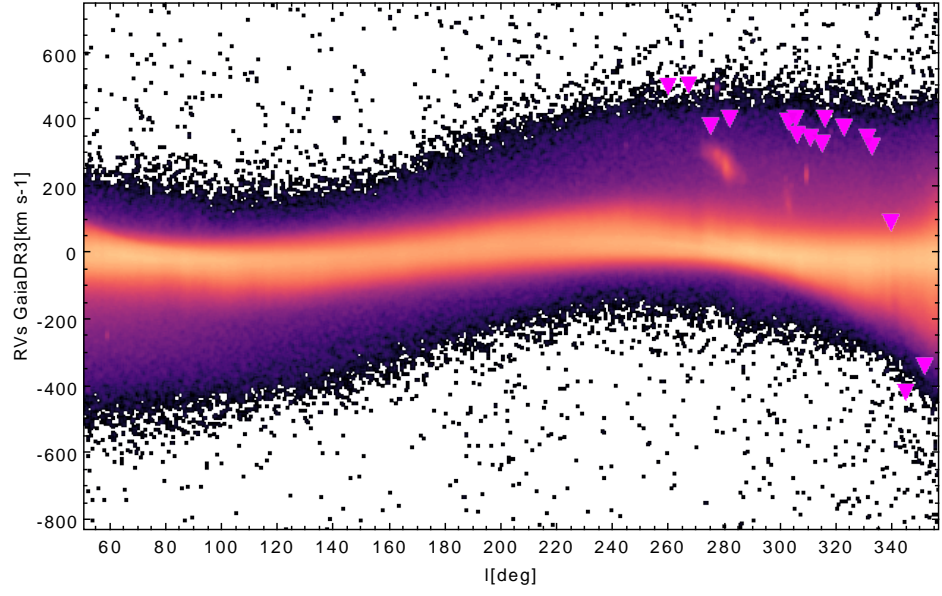}
\caption{$V_r$ for our sample (pink triangles) compared with the Gaia DR3 radial velocities.}
\label{fig:katz2025}
\end{figure}

\section{Carbon, Nitrogen, Oxygen and Carbon isotopic ratio analysis}

\begin{figure}
\centering
\includegraphics[width=0.50\textwidth]{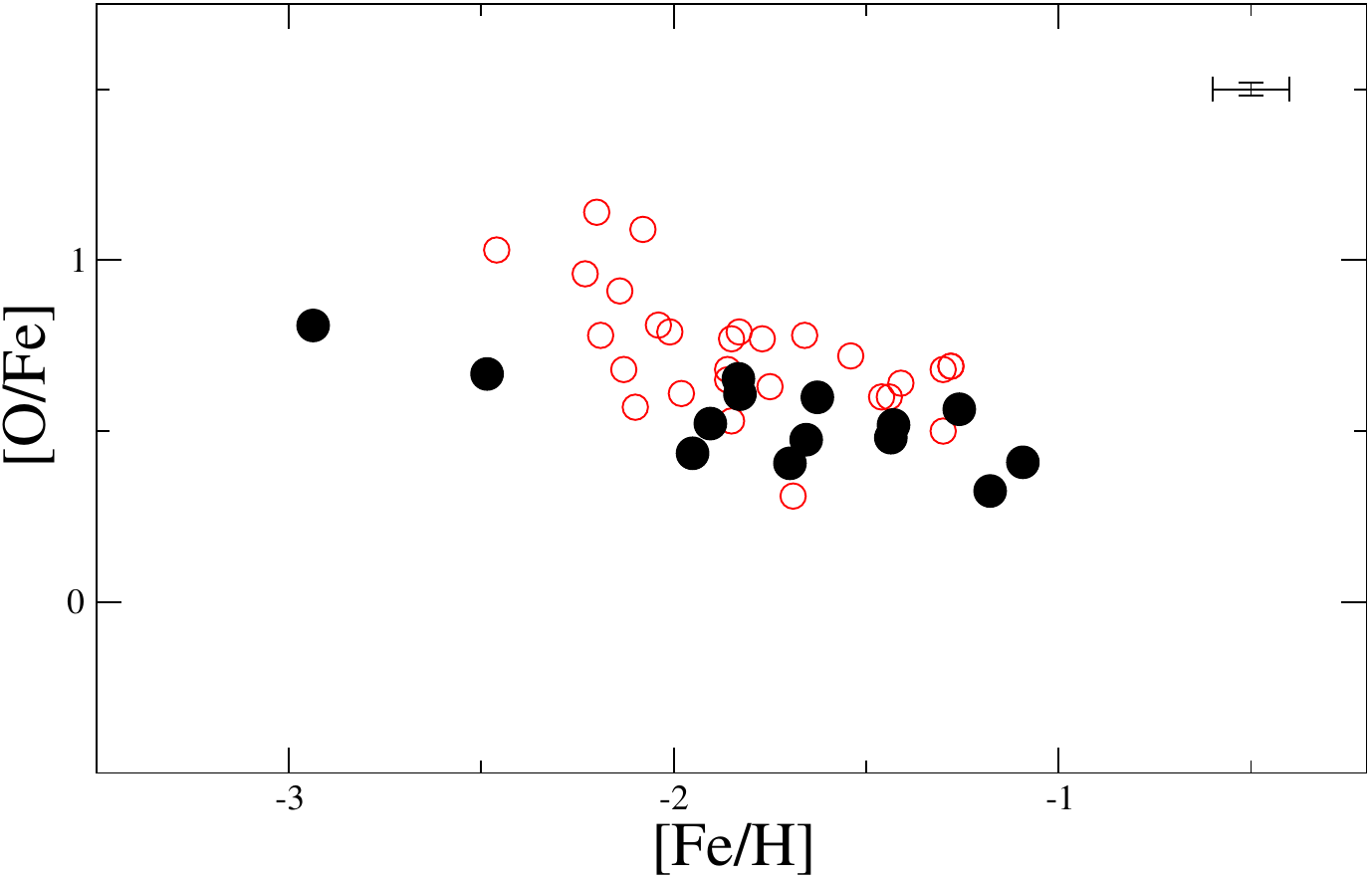}
\caption{Our sample compared with the sample from \cite{Cescutti2022}}
\label{fig:O_abun}
\end{figure}

For the abundance analysis, I used MyGIsFOS to determine the oxygen abundances. Oxygen was measured using the forbidden [O I] lines at 630.0 nm and 636.3 nm. The derived abundances are presented in Figure~\ref{fig:O_abun}, where they are compared with the MINCE sample \citep{Cescutti2022}. Oxygen abundances are plotted against Fe II abundances. Our sample (black symbols) is shown alongside metal-poor stars from the MINCE project (red open symbols).

\newpage
To determine the carbon abundance, I fitted the CH G-band at 430\,nm. The molecular data were adopted from \citet{Masseron2014}. Nitrogen abundances were derived from the NH band at 336\,nm, using the molecular data from \citet{Fernando2018}. The spectral regions hosting the G-band and NH band are crowded with atomic lines, requiring careful fitting of the molecular data.

\begin{figure}
\centering
\includegraphics[width=0.50\textwidth]{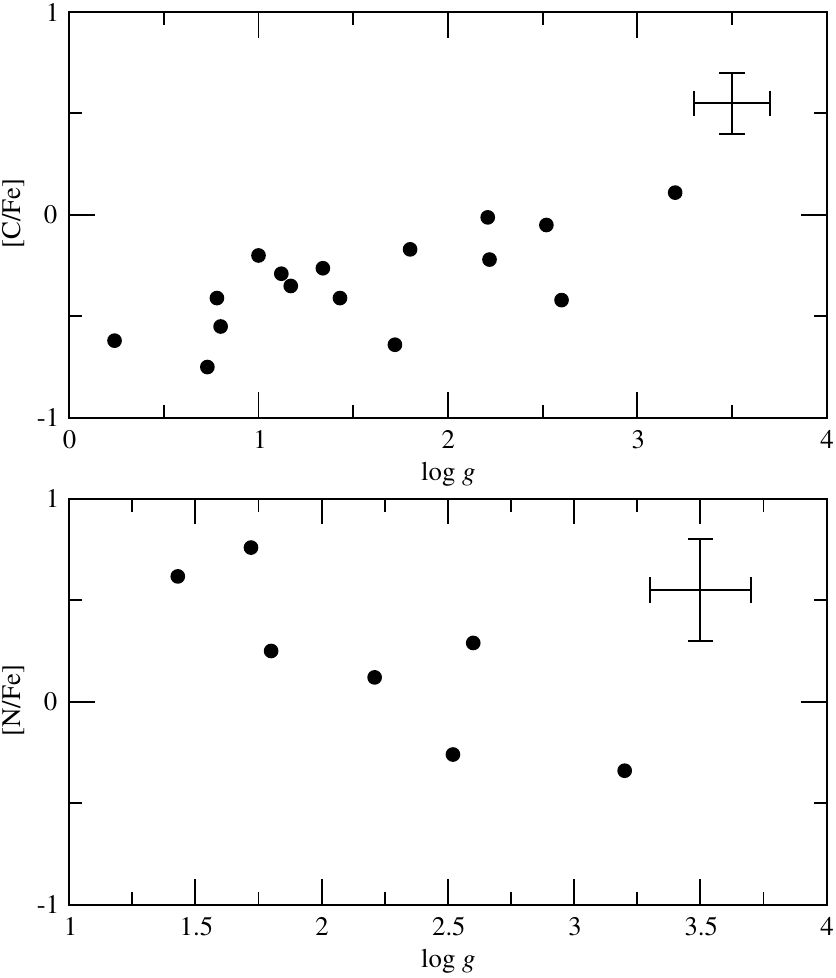}
\caption{Carbon and Nitrogen abundances versus surface gravity.}
\label{fig:C_N_logg}
\end{figure}

Figure~\ref{fig:C_N_logg} shows the behaviour of carbon and nitrogen abundances as a function of surface gravity. As expected, evolved stars tend to show enhanced nitrogen and depleted carbon. This trend reflects the effects of the internal mixing processes.

\begin{figure}
\centering
\includegraphics[width=0.50\textwidth]{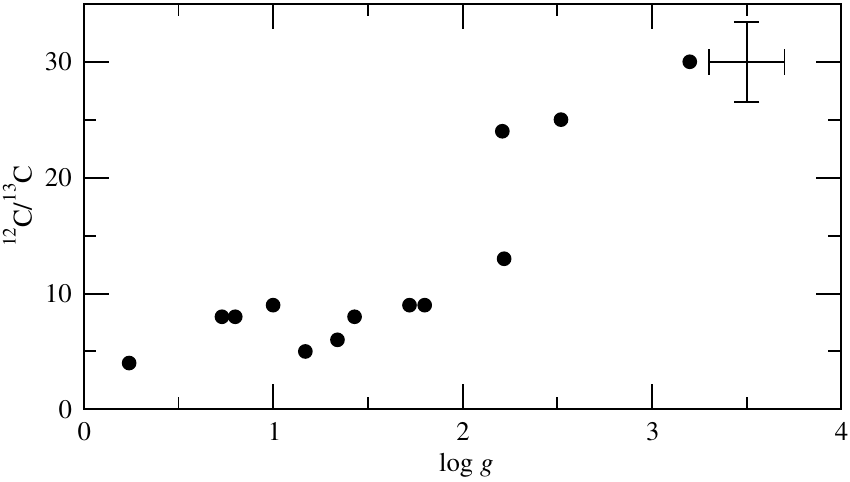}
\caption{The relation between Carbon isotopic ratio and surface gravity}
\label{fig:logg_C12C13}
\end{figure}

I was also able to estimate the $^{12}\mathrm{C}/^{13}\mathrm{C}$ ratio. Most of the stars in the sample show low isotopic ratios ($^{12}\mathrm{C}/^{13}\mathrm{C} < 10$), which is a clear signature of internal mixing. These results are consistent with the atmospheric parameters derived for these stars.
As shown in Figure~\ref{fig:logg_C12C13}, the plot indicates that the more evolved stars tend to have $^{12}\mathrm{C}/^{13}\mathrm{C} < 10$, confirming that these stars have undergone mixing.

We summarise in Table\,\ref{tab:CNO} the abundances we derived for oxygen (from the [OI] forbidden lines), carbon (the G-band), nitrogen (the NH band) and the Carbon isotopic ratio.

\begin{table*}[tb]
\caption{CNO Abundances for the analysed stars.}
\label{tab:CNO}
\small
\centering
\begin{tabular}{lccccccc}
\hline
Star name & $T_{\mathrm{eff}}$ (K) & $\log g$ & [Fe/H] & A(C) & A(N) & A(O)  & 12C/13C \\
\hline
 TYC7274-00734-1  & 5032 & 2.52 & $-1.52$ & 6.86 & 6.050 & 7.81  & 25 \\
 TYC7535-00160-1  & 4649 & 1.43 & $-2.67$ & 5.35 & 5.778 & 6.942  & 8  \\
 TYC8019-00159-1  & 4864 & 1.72 & $-2.51$ & 5.28 & 6.080 & no-Oxy & 9 \\
 TYC7524-0006501  & 4503 & 1.17 & $-2.04$ & 6.04 &    &  7.37  & 5  \\
 C0213360-505024  & 5486 & 2.60 & $-1.42$ & 6.59 & 6.70 & 7.91  &  \\
 C2136142-694908  & 4561 & 1.80 & $-1.19$ & 7.07 & 6.89 & 8.076  & 9 \\
 C1919566-632839  & 4069 & 0.80 & $-1.92$ & 5.96 &     & 7.579 &   $\sim 8$ \\
 C1519196-191359  & 4319 & 1.34 & $-2.08$ & 6.087 &    & 7.468  & 6 \\
 C1032508-244851  & 4970  & 2.21 & $-2.10$ & 6.318 & 5.85 & 7.244  & 27 \\
 C1536201-144228  & 4888  & 2.22 & $-1.96$ & 6.25 &      & 7.58  & 13 \\
 C1302091-323721  & 4223  & 0.78 & $-1.67$ & 6.35 &    & 7.732  &  \\
 C1250398-030748  & 5443  & 3.20 & $-2.15$ & 6.39 & 5.34 & No-Oxy  & $\sim 30$\\
 C1126364-293415  & 4196  & 1.12 & $-1.43$ & 6.71 &    & 8.066  &  \\
 C1012254-203007  & 4520  & 0.73 & $-2.91$ & 4.77 &    & 6.632  &  $\sim 8 $ \\
 C1304064-331913  & 4189  & 1.00 & $-1.68$ & 6.55 &    & 7.849   & 9\\
 C1330590-230955  & 4000  & 0.24 & $-1.85$ & 5.96 &    & 7.540  &  $\sim 4$ \\
 
\hline
\end{tabular}
\end{table*}

\begin{figure}
\centering
\includegraphics[width=0.50\textwidth]{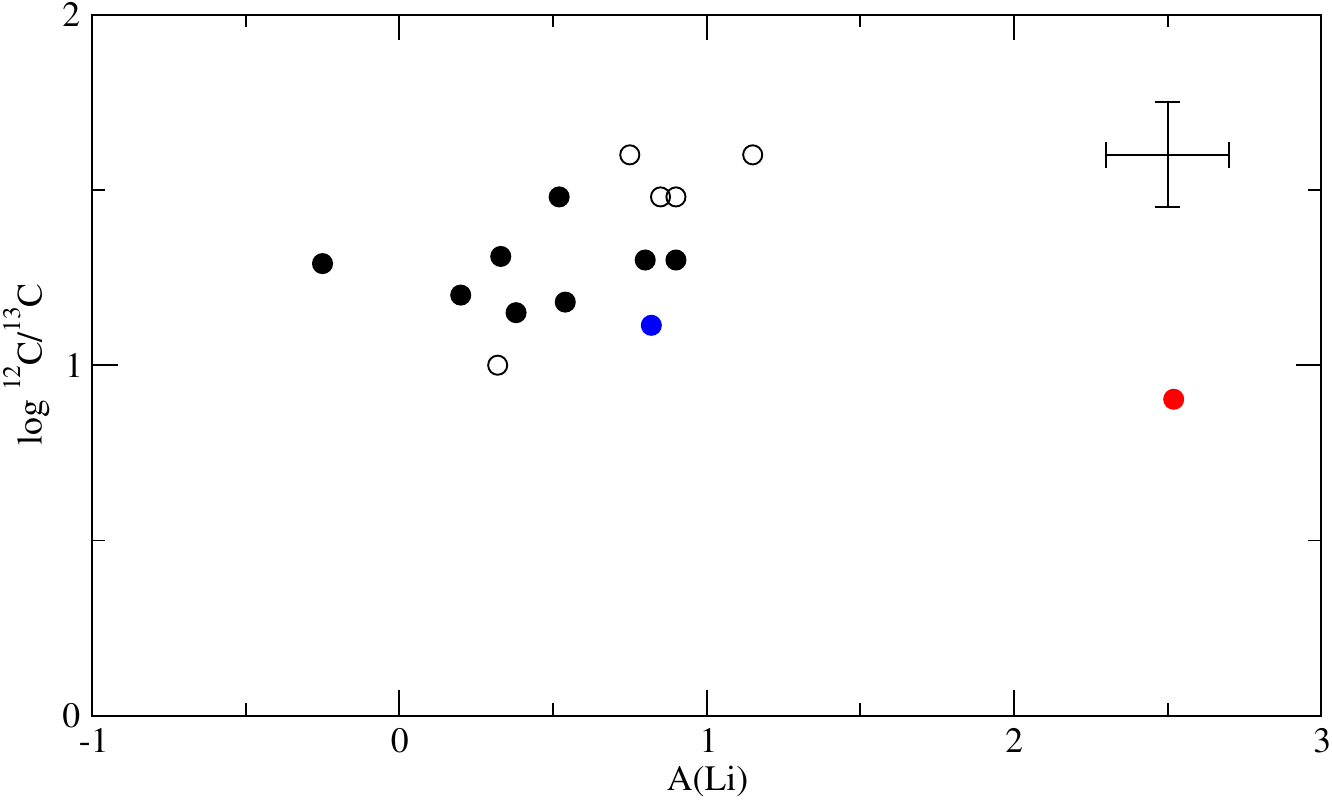}
\caption{Lithium abundances compared with the sample of \citet{Spite2006}. Circles represent the \citet{Spite2006} sample and the open circles indicate upper limits. The red point corresponds to C1012254$-$203007 and the blue one is for the star C1536201$-$144228.}
\label{fig:Li}
\end{figure}

\subsubsection*{Lithium}

Lithium is detected in two stars of the sample. The Li\,I line at 670.8\,nm in C1012254$-$203007 is particularly strong. This fact was previously reported by \citet{Ruchti2011} investigating MIKE/Magellan spectra. They derived A(Li)$=2.52$ under LTE assumptions, which decreases to A(Li)$\sim 2.3$ after NLTE corrections. The spectrum here investigated confirms the strength of this line, in agreement with their results.
A second star, C1536201$-$144228, also shows the Li\,I 670.8\,nm feature, although it is weaker. I derived A(Li)$=0.82$ in 1D-LTE. I compare in Figure~\ref{fig:Li} the lithium abundances and carbon isotopic ratios of these stars with the sample of \citet{Spite2006}.
C1536201$-$144228 (blue dot in the figure) is perfectly consistent with the comparison sample, while C1012254$-$203007 has a much larger Li abundance.  
\newpage

\section*{Acknowledgements}
  We  are grateful to David Katz for his help in using   Gaia radial velocities. 
  
\bibliographystyle{aa}
\bibliography{template}

\end{document}